\documentclass[emulateap5,preprint2]{aastex}

\newcommand\om{\Omega_m}
\newcommand\ob{\Omega_b}
\newcommand\vs{V_{\rm Sloan}}
\newcommand\knl{k_{\rm nl}}

\shorttitle{Probing dark energy using baryonic oscillations}

\shortauthors{Chris Blake \& Karl Glazebrook}

\begin{document}

\title{Probing dark energy using baryonic oscillations in the galaxy
power spectrum as a cosmological ruler}

\author{Chris Blake\altaffilmark{1}}
\affil{School of Physics, University of New South Wales, Sydney, NSW
2052, Australia}
\email{chrisb@phys.unsw.edu.au}
\and
\author{Karl Glazebrook}
\affil{Department of Physics \& Astronomy, Johns Hopkins University,
Baltimore, MD 21218-2686, United States}
\email{kgb@pha.jhu.edu}

%\slugcomment{Submitted to The Astrophysical Journal}

\altaffiltext{1}{Visiting scholar, Department of Physics \& Astronomy,
Johns Hopkins University}

\begin{abstract}
We show that the baryonic oscillations expected in the galaxy power
spectrum may be used as a ``standard cosmological ruler'' to
facilitate accurate measurement of the cosmological equation of state.
Our approach involves a straight-forward measurement of the
oscillation ``wavelength'' in Fourier space, which is fixed by
fundamental linear physics in the early Universe and hence is highly
model-independent.  We quantify the ability of future large-scale
galaxy redshift surveys with mean redshifts $z \approx 1$ and $z
\approx 3$ to delineate the baryonic peaks in the power spectrum, and
derive corresponding constraints on the parameter $w$ describing the
equation of state of the dark energy.  For example, a survey of three
times the Sloan volume at $z \approx 1$ can produce a measurement with
accuracy $\Delta w \approx 0.1$.  We suggest that this method of
measuring the dark energy powerfully complements other probes such as
Type Ia supernovae, and suffers from a different (and arguably less
serious) set of systematic uncertainties.
\end{abstract}

%\keywords{cosmological parameters -- large-scale structure of Universe
%-- surveys}

\section{Introduction}

Measurement of anisotropies in the Cosmic Microwave Background (CMB)
radiation have shown the Universe to be (very close to) spatially
flat.  However, matter only makes up a third of the critical density
-- the dominant contribution to the energy density of the Universe
appears to exist in an unclustered form called ``dark energy''.
Furthermore, observations of distant supernovae have indicated that
the expansion of the Universe has entered a phase of acceleration.
This is explicable if dark energy has recently come to dominate the
dynamics of the Universe and exerts a large negative pressure.  This
situation occurs naturally if Einstein's equations contain a
cosmological constant, but the observed magnitude of the vacuum energy
is wildly inconsistent with the current predictions of quantum
physics.  This has motivated the consideration of a more general dark
energy equation of state, $P = w \, \rho$ (Turner \& White 1997).  An
accelerating universe is produced if $w < -1/3$ and vacuum energy is
described by $w = -1$.  If the dark energy is the product of new
physics, such as quintessence (e.g. Ratra \& Peebles 1988), then the
resulting equation of state varies with redshift and must be treated as
a more general $w(z)$ (e.g. Linder \& Huterer 2003).

Securing more accurate measurements of the dark energy component is of
prime cosmological importance.  A number of methods have been
investigated, most notably the use of Type Ia supernovae as ``standard
candles'' (e.g. Weller \& Albrecht 2002).  Other possible probes
include counts of galaxies (Newman \& Davis 2000) and of clusters
(Haiman et al. 2001), weak gravitational lensing (Cooray \& Huterer
1999), the Alcock-Paczynski test applied to small-scale galaxy
correlations (Ballinger et al. 1996) and use of the CMB (e.g. Douspis
et al. 2003).  In this paper we examine a probe of dark energy that
has not received much attention in the literature (e.g. Lahav 2002),
but which we believe has many advantages: baryonic oscillations in the
galaxy power spectrum.

The CMB power spectrum contains acoustic peaks.  Coupling between
baryons and photons at recombination imprints these ``wiggles'' into
the matter power spectrum on a scale corresponding to the sound
horizon in the early Universe (Peebles \& Yu 1970, Eisenstein \& Hu
1998).  The positions of the peaks and troughs in Fourier space are
calculable from straight-forward linear physics and act like a
``standard cosmological ruler'' (Eisenstein, Hu \& Tegmark 1998).
Therefore a power spectrum analysis of a galaxy redshift survey
containing acoustic oscillations can be used to measure the
cosmological parameters (Eisenstein 2002): the conversion of the
redshift data into real space requires values of the parameters to be
assumed; an incorrect choice leads to a distortion of the power
spectrum and the appearance of the acoustic peaks in the wrong places.
Standard ruler techniques for deducing the cosmic parameters have been
proposed many times before.  In the Alcock-Paczynski test (Alcock \&
Paczynski 1979) a spherical system is distorted into an ellipsoid in
the wrong cosmological model.  Other examples include the detection of
a characteristic clustering scale within the 2dF quasar survey
(Roukema, Mamon \& Bajtlik 2002) and the use of the angular power
spectrum of dark matter haloes (Cooray et al. 2002).  Our study is
focussed on the ability of the standard ruler provided by the baryonic
peaks to measure the dark energy.

Mapping the acoustic peaks in the galaxy power spectrum at high
precision, matching those already measured in the CMB power spectrum,
would provide a spectacular confirmation of the standard cosmological
model in which mass overdensities grow from the seeds of CMB
fluctuations.  The sharp features of the baryonic oscillations
represent a powerful and precise observational test of the current
cosmological paradigm (``$\Lambda$-CDM'').  Accurate delineation of
the baryonic oscillations lies beyond the current state-of-the-art
galaxy redshift surveys, the 2dF Galaxy Redshift Survey (Colless et
al. 2001) and Sloan Digital Sky Survey (York et al. 2000).  Moreover,
it is advantageous to place the survey volume at higher redshift
(Eisenstein 2002), where the linear regime of galaxy clustering
extends to smaller physical scales (unveiling acoustic oscillations to
higher spatial frequencies).

In this initial study we present a simple ``proof-of-concept'' that
acoustic oscillations measured at high redshift may be used to
accurately determine the dark energy parameter $w$, restricting our
attention to cosmological models with constant (redshift-independent)
$w$.  We perform simulations to quantify the scale of redshift survey
required to recover the wiggles and determine the resulting
constraints on $w$.  We suggest that measurement of the dark energy
should be a prime motivation for the next-generation galaxy redshift
survey.

\section{Measuring baryonic oscillations}
\label{seclammeas}

In this section we investigate the scale of redshift survey required
to measure the acoustic oscillations in the galaxy power spectrum with
a specified accuracy.

\subsection{Assumptions and approximations}

We modelled the baryonic oscillations using the transfer function
fitting formulae of Eisenstein \& Hu (1998).  The matter power
spectrum $P(k)$ is deduced from the transfer function $T(k)$ using
$P(k) = A \, k^n \, T^2(k)$; we took the primordial spectral slope to
be $n = 1$, as (approximately) suggested by inflationary models, and
fixed the normalization $A$ such that $\sigma_8 = 1$.  The transfer
function model is specified by assigning values to the matter density
$\om$, the baryon density $\ob$ and Hubble's constant $h = H_0/$(100
km s$^{-1}$ Mpc$^{-1}$).  We assumed that it does not depend on the
dark energy equation of state: the energy density of dark energy,
$\rho_x$, relative to that of matter, $\rho_m$, scales with redshift
as $\rho_x/\rho_m \propto (1+z)^{3w}$ and is a negligible contributor
to physics before recombination (as $w < -1/3$).  Of course more
complex dark energy models with varying $w$ may lead to significant
effects at the epoch of recombination; we do not consider these here,
but point out that the CMB is well described by simple
matter-dominated models (Spergel et al. 2003).  In this section we
assume $\om = 0.3$, $\ob/\om = 0.15$, $h = 0.7$ and $w = -1$ (i.e. a
cosmological constant).  Our assumed value for $\ob$ is consistent
with the latest CMB results (Spergel et al. 2003), constraints from
Big Bang nucleosynthesis theory (e.g. O'Meara et al. 2001), and the
shape of the galaxy power spectrum measured by the 2dF Galaxy Redshift
Survey (Percival et al. 2001).  We assume throughout this study that
the Universe is flat, $\Omega_{\rm tot} = 1$.

\begin{figure}
\begin{center}
\includegraphics[width=7cm,angle=0]{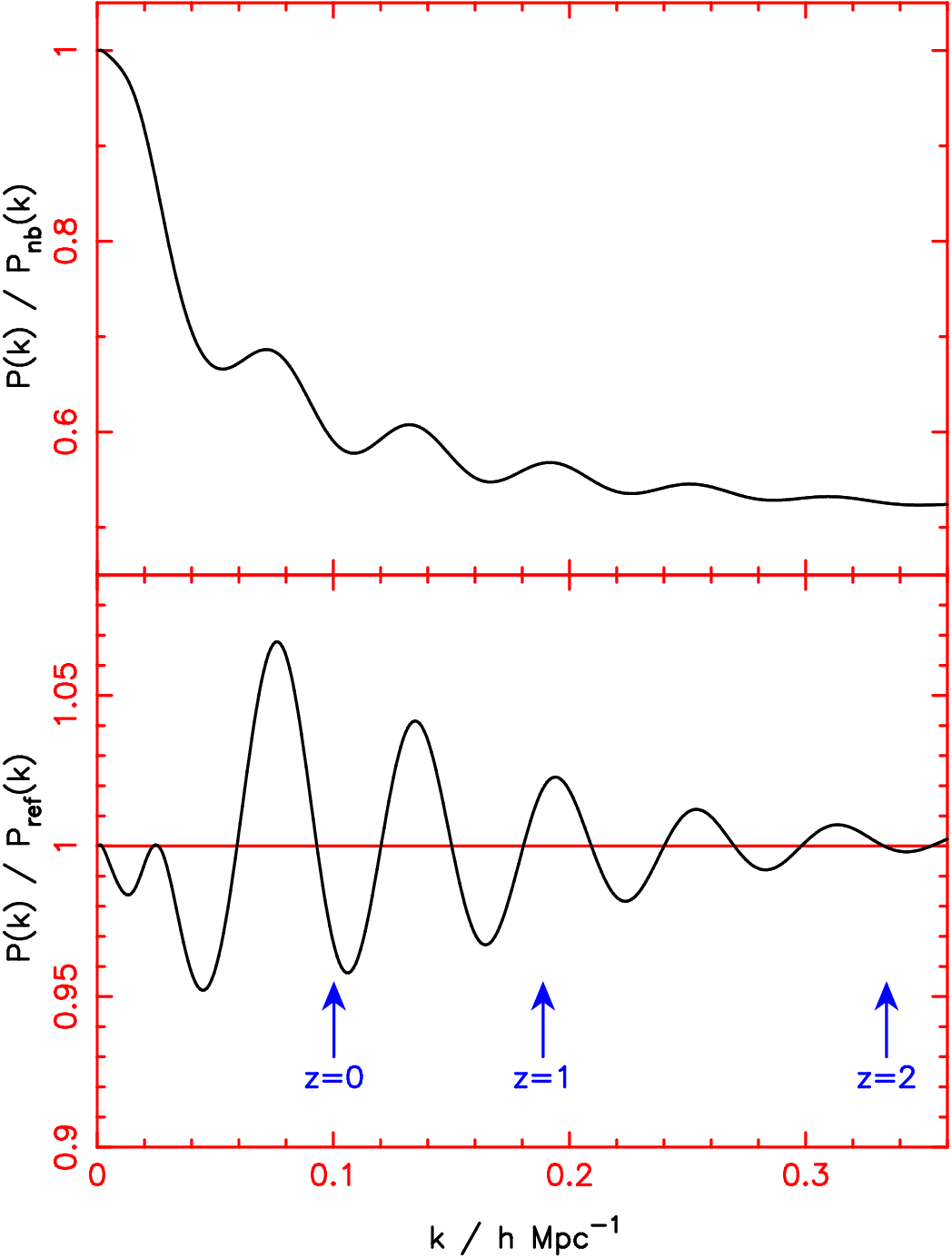}
\end{center}
\caption{The model power spectrum of Eisenstein \& Hu (1998) for
cosmological parameters $\om = 0.3$, $\ob/\om = 0.15$, $h = 0.7$.  For
this choice of parameters, the sound horizon $s \approx 105 \, h^{-1}$
Mpc and the wiggle wavescale $k_A \approx 0.0601 \, h$ Mpc$^{-1}$.  In
the upper panel we divide the model $P(k)$ by the corresponding
zero-baryon model, $\om = 0.3$, $\ob = 0$, $h = 0.7$.  Replacing cold
dark matter with baryons produces acoustic oscillations and an overall
suppression of power.  In the lower panel we divide the model $P(k)$
by a smooth fit to the overall shape of the spectrum.  The arrows
indicate the approximate position of the linear/non-linear transition
at different redshifts, estimated in the following way.  At $z=0$, we
conservatively defined the linear regime by $k < \knl = 0.1 \, h$
Mpc$^{-1}$.  From the model $P(k)$ we computed the variance of mass
fluctuations $\sigma^2 (R)$ inside a sphere of radius $R$, where $R =
\pi/2\knl$ (i.e. half a fluctuation wavelength, or whole wavecrest,
coincides with the diameter $2R$).  For $\knl = 0.1 \, h$ Mpc$^{-1}$
we found $\sigma^2 (\pi/2\knl) = 0.35$, and then applied this
criterion to fix the linear/non-linear transition at other redshifts.
At higher $z$, the amplitude of $P(k)$ is reduced by the growth
factor, $P(k) \rightarrow P(k) \, D_1(z)^2$.  At $z = 1$, for example,
$\sigma^2 (\pi/2k) = 0.35$ for $k = 0.19 \, h$ Mpc$^{-1}$ $= \knl (z =
1)$.  We fixed the overall amplitude of $P(k)$ such that $\sigma^2 (8
\, h^{-1} \, \rm{Mpc}) = 1$.}
\label{figpkmod}
\end{figure}

The baryonic oscillations can be conveniently displayed by dividing
the model $P(k)$ by a smooth reference spectrum containing no wiggles
(lower panel of Figure \ref{figpkmod}; the reference spectrum was
obtained from the fitting formulae of Eisenstein \& Hu (1998) together
with the transfer function).  The result is well-approximated by a
slowly-decaying sinusoidal function whose peaks are harmonics of the
sound horizon scale: approximate formulae for their positions in
Fourier space are given by Eisenstein \& Hu (1998).  The amplitude of
the wiggles increases with the baryon density $\ob$.  The key quantity
we wish to observe is the ``wavelength'' of the acoustic oscillations
in $k$-space, which we denote as $k_A$, and which is related to the
sound horizon at recombination by $k_A = 2\pi/s$.

\placefigure{figpkmod}

If dark energy is neglected at high redshift, the co-moving sound
horizon size at last scattering, $s$, is given by (Hu \& Sugiyama
1995, Cornish 2001):
\begin{equation}
 s= \frac{1}{H_0 \, \om^{1/2}} \int_0^{a_r} \frac{c_s}{(a + a_{\rm
 eq})^{1/2}} \, da
\label{eqhorizon}
\end{equation}
where $a_r$, $a_{\rm eq}$ are the values of the scale factor $a =
1/(1+z)$ at recombination and matter-radiation equality respectively.
$c_s$ is the speed of sound and is approximately $c/\sqrt{3}$ over the
interval of integration.  The value of $s$ is of order 100 $h^{-1}$
Mpc.

Thus the {\it theoretical} value of $k_A$ is set by fundamental CMB
physics and depends strongly on $\om$, weakly on $\ob$ and negligibly
on dark energy.  $k_A$ is our ``standard measuring rod''.  In a
redshift survey at intermediate redshift $z$, the {\it apparent} size
of the oscillation wavescale $k_A$ will depend on the cosmological
geometry, now including the effects of dark energy.  As we will see in
Section \ref{secwmeas}, the effects of assuming an incorrect world
model (e.g. incorrect $w$) would include a distortion of the measured
value of $k_A$.  (There will be different effects parallel and
perpendicular to the line-of-sight, but these are averaged out in our
approach).  To zeroth order, the precision with which we can
empirically measure $k_A$ tells us how accurately we can measure the
geometrical distance to the redshift $z$ and hence how accurately we
can measure the equation of state.

We have assumed that the power spectra of galaxies and of matter are
related by a linear bias factor $b$ (where $P_{\rm gal} = b^2 \,
P_{\rm mat}$).  Provided our simulated surveys are dominated by sample
variance (see Section \ref{secsimpcalc}), the value of $b$ is
unimportant: the fractional error bars $\sigma_P/P$ (i.e. the
appearance of Figures \ref{figpkmod} and \ref{figpkreal}) do not
change if $P(k)$ is scaled by a constant factor (see equation
\ref{eqerr}).  However, increasing the value of $b$ (hence increasing
$P_{\rm gal}$) reduces the fractional error due to shot noise, and
hence reduces the number of objects required to render shot noise
negligible, by a factor $b^2$.  The assumption of linear bias is
simplistic and we plan to include more complicated
(redshift-dependent) biassing schemes in a future study.

We scaled the power spectrum to redshift $z$ using the appropriate
growth factor, $P(k,z) = P(k,0) \, D_1(z)^2$.  Throughout this paper
we use the approximation of Carroll, Press \& Turner (1992) for
$D_1(z)$.  This formula is only valid for $w = -1$, but is a
satisfactory estimate for models with $w \ne -1$, given that we only
treat small departures from $w = -1$ and that we divide out the
overall shape of $P(k)$ in the analysis procedure.

These model power spectra are only valid in the linear regime of
structure formation; on smaller scales the non-linear growth of
structure washes out the baryonic oscillations.  At redshift zero, the
linear regime only extends to $k \lesssim 0.1 \, h$ Mpc$^{-1}$ (see
Meiksin, White \& Peacock 1999, Percival et al. 2001) and perhaps only
the first major peak shown in Figure \ref{figpkmod} is preserved
(i.e. the peak at $k \approx 0.075 \, h$ Mpc$^{-1}$; there is a small
peak predicted at $k \approx 0.025 \, h$ Mpc$^{-1}$ which is
undetectable due to cosmic variance).  At higher redshifts ($z \gtrsim
1$) the linear regime extends to much smaller scales, unveiling more
acoustic peaks.  Moreover, a high-redshift study is necessary to probe
dark energy effectively: the conversion of redshifts to spatial scales
only depends on the Hubble constant $H_0$ at low $z$ and cannot
distinguish between different values of $w$.

A galaxy power spectrum deduced from a redshift survey is subject to
redshift-space distortions owing to the peculiar velocities galaxies
possess on top of the bulk Hubble flow.  There are two effects (Kaiser
1987): the coherent infall of galaxies into concentrations of mass
(which boosts power on all scales by a constant factor) and the
incoherent velocities of galaxies in the central regions of clusters
(which damps power on small scales).  We make no attempt to simulate
these effects.  Redshift-space distortions will only produce smooth
changes in the shape of the angle-averaged $P(k)$, not sharp features
such as acoustic peaks, and we note again that the fractional error
bars $\sigma_P/P$ do not depend on the absolute value of $P(k)$
(equation \ref{eqerr}).  In fact, we explicitly divide out the shape
using the smooth reference spectrum.  Moreover, incoherent velocities
do not have an important effect on the large scales probed by acoustic
oscillations.  These velocities can be modelled by a 1D pairwise
dispersion $\sigma_p$, and become important on scales $k$ where $k \,
\sigma_p \gtrsim H_0$.  At redshift zero, $\sigma_p \approx 300$ km
s$^{-1}$ and power is hence damped on scales $k \gtrsim 0.3 \, h$
Mpc$^{-1}$.  Furthermore, simulations indicate that the value of
$\sigma_p$ drops by a factor $\sim 2$ between $z = 0$ and $z = 2$
(Magira, Jing \& Suto 2000).

Some models of galaxy clustering (Peacock \& Smith 2000, Seljak 2000)
predict that the baryonic oscillations in the power spectrum are
diluted by a wiggle-free halo contribution (which amounts to a
non-local recipe for bias, in which the probability of finding a
galaxy is not a simple function of the local mass density).  In our
initial treatment we assume a standard linear bias factor.  However,
this should not modify our results to first order: at $z = 0$, the
halo contribution to the matter power spectrum becomes equal to the
linear contribution at $k \approx 0.4 \, h$ Mpc$^{-1}$ (Peacock \&
Smith 2000, Figure 2), whereas most of our high-redshift constraining
power (all of it in the case of $z \sim 1$) originates from $k < 0.2
\, h$ Mpc$^{-1}$.  At higher redshifts the transition in these halo
clustering models scales to higher $k$ as described in Smith et
al. (2002, Figure 15).

\subsection{Back-of-the-envelope calculation}
\label{secsimpcalc}

A simple calculation demonstrates that to delineate the oscillations
in the galaxy power spectrum accurately, we need to survey a
cosmological volume greater than that of the local Sloan Digital Sky
Survey (which we take as a uniform cone of $10^4$ deg$^2$ to $z =
0.2$, i.e. $\vs = 2 \times 10^8 \, h^{-3}$ Mpc$^3$).  There are two
sources of statistical error in a power spectrum measurement: {\it
sample variance} (the number of independent wavelengths $2\pi/k$ of a
given fluctuation that can fit into the survey volume $V$) and {\it
shot noise} (the imperfect sampling of fluctuations by the finite
number of galaxies $N$).

The error due to sample variance on a power spectrum measurement,
averaged over a radial bin in $k$-space of width $\Delta k$, is
\begin{equation}
\left( \frac{\sigma_P}{P} \right)^2 = 2 \times
\frac{(2\pi)^3}{V} \times \frac{1}{4\pi k^2 \Delta k}
\label{eqerr}
\end{equation}
(e.g. Peacock \& West 1992).  This expression is derived from the
density of states and the volume of $k$-space enclosed by the radial
bin.  The initial factor of $2$ is due to the fact that the density
field is real rather than complex, thus only half the modes are
independent.  (For a realistic survey geometry equation \ref{eqerr} is
only an approximation, as neighbouring $k$-modes are correlated).
Figure \ref{figpkmod} indicates that for a statistically significant
measurement of the acoustic peak at $k \approx 0.135 \, h$ Mpc$^{-1}$
we require $\sigma_P/P \lesssim 0.02$ as the fractional height of this
peak is about 4 per cent.  Thus taking $\Delta k = 0.015 \, h$
Mpc$^{-1}$ ($= k_A/4$) we find that $V \gtrsim 1.8 \, \vs$.

The number of galaxies needed to render shot noise negligible in this
case is $N \gg 1/P \approx 1 \times 10^5$ (assuming $b = 1$ and using
the input model power spectrum to read off $P \times V \approx 5
\times 10^3 \, h^{-3}$ Mpc$^3$ at $k \approx 0.135 \, h$ Mpc$^{-1}$).

\subsection{Simulations}

We constructed detailed simulations to determine the scale of survey
required to measure the oscillation wavescale $k_A$ with a given
accuracy.  When measuring a power spectrum from a realistic survey,
the quantity we derive is actually the power spectrum convolved with
the survey window function in $k$-space, $W(k)$, the Fourier transform
of the window function in real space.  To detect baryonic oscillations
it is important that $W(k)$ is compact (non-zero only for $k \ll
k_A$), otherwise the smoothing effect of the convolution seriously
reduces the amplitude of the oscillations.  This restricts us to
relatively simple survey geometries.  (Any remaining convolution has
no effect on the observed oscillation wavescale $k_A$).

We considered two different potential surveys: the first directed at
redshift range $0.5 < z < 1.3$ and the second probing the range $2.5 <
z < 3.5$.  These ranges are reasonable possibilities for future
large-scale surveys.  The lower-redshift survey could target either
luminous star-forming galaxies via 3727\AA$\,$ [OII] emission, or
luminous elliptical galaxies using absorption features such as the
4000\AA$\,$ continuum break.  In either case, $z = 1.3$ is a
reasonable limit where these spectral features move out of optical
wavebands.  For the higher redshift survey, the Lyman Break selection
technique operates efficiently for $2.5 < z < 3.5$ as the 912\AA$\,$
break shifts through the optical $U$-band.  Both these redshift
regimes have already been explored by small surveys of hundreds of
galaxies (e.g. Le Fevre et al. 1995, Steidel et al. 1998) and it is
now becoming feasible to extend these efforts to far larger surveys.

For each redshift range $0.5 < z < 1.3$ and $2.5 < z < 3.5$ we defined
a survey volume $V$ by an angular patch on the sky of size $\theta
\times \theta$.  We varied the volume $V$ by changing $\theta$ such
that $0 < V < 6 \, \vs$ in each case ($\theta_{\rm max} = 24.9^\circ$
and $17.3^\circ$ respectively).  For the $z \sim 1$ survey we
restricted the power spectrum measurements to the Fourier range $k <
0.2 \, h$ Mpc$^{-1}$, which encompasses the first two detectable
acoustic peaks.  Non-linearities and redshift-space distortions are
likely to become important for $k > 0.2 \, h$ Mpc$^{-1}$, diluting the
baryonic features.  Our estimate of the linear/non-linear transition
scale $\knl(z=1) \approx 0.2 \, h$ Mpc$^{-1}$ is based on using the
model power spectrum to calculate the dimensionless variance of mass
fluctuations, $\sigma^2(R)$, inside a sphere of co-moving radius $R$
equivalent to a fluctuation of wavelength $2\pi/k$.  We defined
equivalent scales by $R = \pi/2k$, matching a half-wavelength (a full
wavecrest) to the diameter $2R$ of the spheres.  A conservative
estimate of $\knl(z=0)$ is $0.1 \, h^{-1}$ Mpc (Meiksin, White \&
Peacock 1999) which corresponds to $\sigma^2(\pi/2\knl) = 0.35$.  At
$z = 1$, this same value of $\sigma^2$ is produced if $\knl \approx
0.2 \, h$ Mpc$^{-1}$ (see Figure \ref{figpkmod}).  For the $z \sim 3$
survey we extended the measured power spectrum range to $k < 0.3 \, h$
Mpc$^{-1}$, which contains effectively all the visible acoustic peaks.
The linear/non-linear transition scale at $z = 3$ is $\knl \approx 0.5
\, h$ Mpc$^{-1}$.

We generated many different Gaussian realizations of $N$ galaxies from
the model $P(k)$ within the survey volumes $V$, using Fast Fourier
Transform techniques.  For simplicity we assumed a constant selection
function across the survey volume (i.e. a constant galaxy number
density).  A more realistic flux-limited survey will detect galaxies
with a varying redshift distribution $N(z)$, but optimal weighting
techniques for deriving $P(k)$ exist in this case (Feldman, Kaiser \&
Peacock 1994).  We also neglected the evolution of $P(k)$ with
redshift over the survey depth, fixing the model power spectrum at the
mean redshift of the survey (i.e. $z_{\rm eff} = 0.9$ or $z_{\rm eff}
= 3$).  This approximation has a negligible effect on the results as
we are only interested in the relative positions of the peaks and
troughs in $P(k)$, which are independent of redshift.

For the $z \sim 1$ survey we assumed a linear bias factor $b = 1$ for
the galaxies.  The potential $z \sim 3$ survey will most likely be
targetted at Lyman Break galaxies, which are known to be strongly
biased tracers of mass (Steidel et al. 1998); in this case we assumed
a linear bias parameter $b = 3$.  As noted above, the main effect of
this assumption is to reduce by a factor $b^2$ the number of galaxies
$N$ required to render shot noise negligible.

We measured the (noisy) power spectrum of each Gaussian realization
using the standard method (see e.g. Hoyle et al. 2002).  The
measurements for the set of realizations allow the statistical error
in each spatial frequency bin to be derived, and the ensemble
incorporates any correlations that exist between adjacent bins.  For
each $(N,V)$ simulation we generated 400 realizations.  For each
realization we divided the measured power spectrum by the smooth
reference spectrum and fitted a decaying sinusoidal function to the
result, deducing a best-fit oscillation wavescale $k_A$ in $k$-space.
We permitted the overall amplitude $A$ of the fitted sinusoidal
function to vary but fixed the decay rate at an empirically-estimated
value, so that we were simply fitting the two-parameter function:
\begin{equation}
\frac{P(k)}{P_{\rm ref}} = 1 + A \, k \, \exp{\left[ - \left(
\frac{k}{0.1 \, h \, {\rm Mpc}^{-1}} \right)^{1.4} \right]} \,
\sin{\left( \frac{2 \pi k}{k_A} \right)}
\label{eqfit}
\end{equation}
The power of $1.4$ in the decay term originates from the Silk damping
fitting formula presented in Eisenstein \& Hu (1998, equation 21).
Varying the decay length as well as the amplitude was found not to
have a significant effect on the fitted values of $k_A$.  For small
$k$ there is a phase shift in the sinusoidal term (Eisenstein \& Hu
1998, equation 22) but this only affects the acoustic peak at $k
\approx 0.025 \, h$ Mpc$^{-1}$ (see Figure \ref{figpkmod}) which is
not measurable due to sample variance.

\begin{figure}[t]
\includegraphics[width=7cm,angle=0]{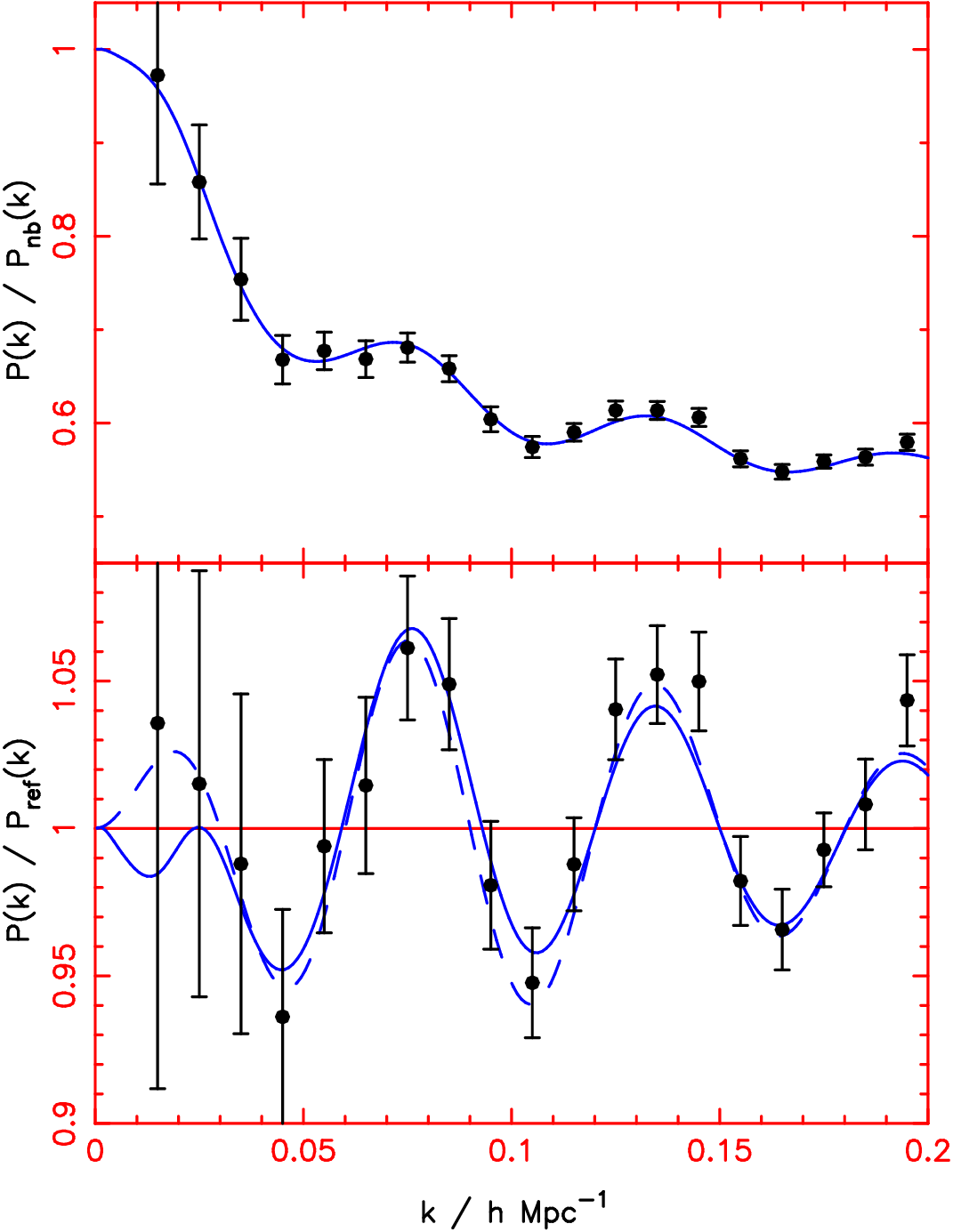}
\caption{Power spectrum measurement for a simulated survey of $N = 2
\times 10^6$ galaxies over a volume $V = 6 \, \vs$ at redshift $z \sim
1$.  The power spectrum is divided by the zero-baryon model in the
upper panel and the smooth reference spectrum in the lower panel.  The
solid line is the input (unconvolved) model power spectrum
(i.e. Figure \ref{figpkmod}).  The dashed line in the lower panel is
the best fit to the data of a simple decaying sinusoidal function
(equation \ref{eqfit}).  Points are plotted at intervals of $\Delta k
= 0.01 \, h$ Mpc$^{-1}$, which are approximately uncorrelated (see
Meiksin, White \& Peacock 1999).}
\label{figpkreal}
\end{figure}

For illustration, Figure \ref{figpkreal} displays the power spectrum
measurement and the best fit of equation \ref{eqfit} for the first
realization of the case $z \sim 1$, $V = 6 \, \vs$, $N = 2 \times
10^6$.

\placefigure{figpkreal}

\begin{figure}[t]
\includegraphics[width=5cm,angle=-90]{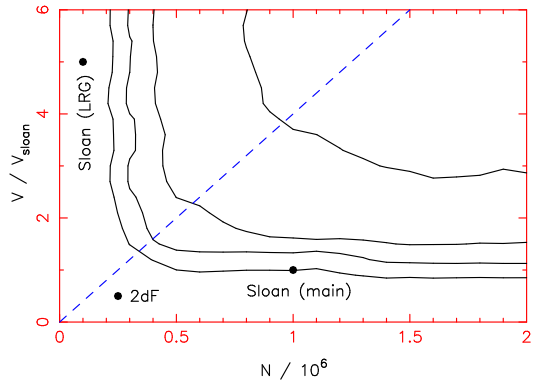}
\caption{Fractional accuracy $\Delta k_A/k_A$ with which the wavescale
of the baryonic oscillations in $k$-space can be measured at redshift
$z \sim 1$, as a function of the number of galaxies $N$ (as a fraction
of $10^6$) and the survey volume $V$ (as a fraction of the Sloan
volume $\vs = 2 \times 10^8 \, h^{-3}$ Mpc$^3$).  Contours are shown
corresponding to (beginning in the bottom left-hand corner) $\Delta
k_A/k_A = 10\%$, $5\%$, $3\%$ and $2\%$.  The positions of the 2dF and
Sloan surveys are marked on the plot for comparison.  This does not
accurately represent their precision in measuring $k_A$, because the
linear regime at redshift zero extends only to $k \approx 0.1 \, h$
Mpc$^{-1}$ (whereas the simulations assume $\knl = 0.2 \, h$
Mpc$^{-1}$).  Also, the Sloan (LRG) sample will possess a higher
linear bias factor $b \approx 1.6$ (whereas the simulations assume $b
= 1$), which enhances the ``effective'' value of $N$ by a factor
$b^2$.  Hence the Sloan LRG sample may measure the position of the
acoustic peak at $k \approx 0.075 \, h$ Mpc$^{-1}$ to an accuracy of
$\sim 5\%$ (Eisenstein, Hu \& Tegmark 1998).  The diagonal dashed line
indicates the most efficient observational strategies: fewer galaxies
will result in shot noise domination, and more galaxies will be
``wasted''.  The dashed line corresponds to a surface density $\approx
2400$ galaxies deg$^{-2}$ in this case.}
\label{figdeltalamz1}
\end{figure}

\begin{figure}[t]
\includegraphics[width=5cm,angle=-90]{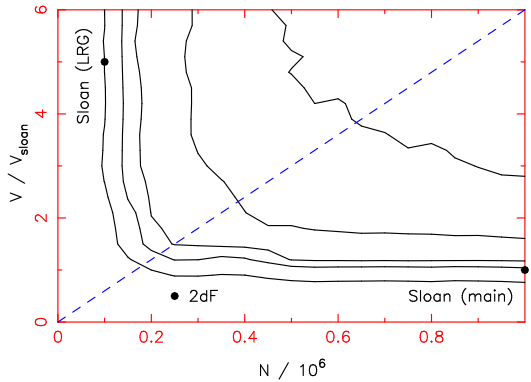}
\caption{The same plot as Figure \ref{figdeltalamz1} for simulated
surveys at redshift $z \sim 3$.  Note the different scale on the
$N-$axis.  Contours are shown corresponding to (beginning in the
bottom left-hand corner) $\Delta k_A/k_A = 10\%$, $5\%$, $3\%$, $2\%$
and $1.5\%$.  The diagonal dashed line corresponds to a surface
density $\approx 3400$ galaxies deg$^{-2}$.}
\label{figdeltalamz3}
\end{figure}

The distribution of fitted oscillation scales over the different
realizations allowed us to describe the accuracy of the measurement by
a quantity $\Delta k_A/k_A$, where $\Delta k_A$ is half the range
enclosed by the $16^{\rm th}$ and $84^{\rm th}$ percentiles of the
distribution (i.e. $k_A \pm \Delta k_A$ defines the 68 per cent
confidence region).  Figures \ref{figdeltalamz1} and
\ref{figdeltalamz3} plot contours of the accuracy $\Delta k_A/k_A$ in
the parameter space of $V$ and $N$ for the two potential surveys at $z
\sim 1$ and $z \sim 3$.  The contours running parallel to the $V-$axis
at small $N$ indicate dominance by shot noise; the contours running
parallel to the $N-$axis at large $N$ represent dominance by sample
variance.  The diagonal dashed lines shown in Figures
\ref{figdeltalamz1} and \ref{figdeltalamz3} provide a reasonably good
representation of the most efficient observational strategies.

\placefigure{figdeltalamz1}

\placefigure{figdeltalamz3}

In order to measure the oscillation wavescale $k_A$ with a precision
of 2 per cent with negligible shot noise, we require either $\sim
10^6$ galaxies covering $\sim 400$ deg$^2$ at $z \sim 1$, or $\sim 5
\times 10^5$ objects over $\sim 150$ deg$^2$ at $z \sim 3$.  The
smaller number of galaxies needed at $z \sim 3$ is due to the larger
bias factor $b = 3$ which boosts the amplitude of the density
fluctuations, more than compensating for the smaller growth factor
$D_1(z)$.  In addition, a given angular field covers more volume at $z
\sim 3$ than at $z \sim 1$.  However, Lyman Break galaxies will
require longer exposure times.  The brightest $z \sim 3$ objects of
Steidel et al. (1998) typically required $\sim 2$ hours of integration
on a 10m telescope to acquire a secure redshift for apparent magnitude
$R \sim 24$.  An $L^*$ galaxy at $z \sim 1$ has brightness $R \sim 23$
and should only require $\sim 30$ minutes exposure.

\section{Measuring dark energy with baryonic oscillations}
\label{secwmeas}

In this section we investigate how accurately a detailed measurement
of the baryonic oscillations can constrain the dark energy parameter
$w$, which we assume does not vary with redshift.  In order to measure
$P(k)$ from a galaxy redshift survey we must convert redshifts to
co-moving co-ordinates (in $h^{-1}$ Mpc), assuming values for $\om$
and $w$.  The expected wiggle wavescale $k_A$ is determined by the
sound horizon before recombination; it is a function of $\om$, $\ob$
and $h$ (see equation \ref{eqhorizon} or the fitting formula of
Eisenstein \& Hu 1998, equation 26).  Incorrect cosmological
parameters will lead to a distortion of the measured power spectrum so
that the deduced value of $k_A$ is not consistent with the theoretical
expectation.

For the purposes of this initial study we assumed that the current
values of the matter density and Hubble's constant are known
precisely, $\om = 0.3$ and $h = 0.7$.  We considered a range of values
for the baryon density $\ob$.  To create a concrete example, we
assumed a fiducial value $w = -1$ for the dark energy and investigated
how accurately we could recover this value from the simulated surveys.

\subsection{Back-of-the-envelope calculation}
\label{secwintro}

For a flat geometry, the fundamental relation between co-moving
distance $x$ and redshift $z$ can be written in the form
\begin{equation}
\frac{dx}{dz} = \frac{c}{H_0 \, \om^{1/2}} \, \frac{1}{\sqrt{(1+z)^3 +
(\om^{-1}-1) (1+z)^{3(1+w)}}}
\label{eqcosmo}
\end{equation}
For values of $w$ near $-1$, the second term inside the $\sqrt{\ }$ is
small for $z \gtrsim 1$ and the zeroth-order dependence is $x \propto
H_0^{-1} \om ^{-1/2}$.  This conveniently cancels the zeroth-order
dependence of the sound horizon scale on $\om$ and $H_0$ in equation
\ref{eqhorizon}.  Thus our cosmological test has reduced sensitivity
to uncertainties in $\om$ and $H_0$.  We quantify this in Section
\ref{secdiscuss} below.

\begin{figure}[t]
\includegraphics[width=5cm,angle=-90]{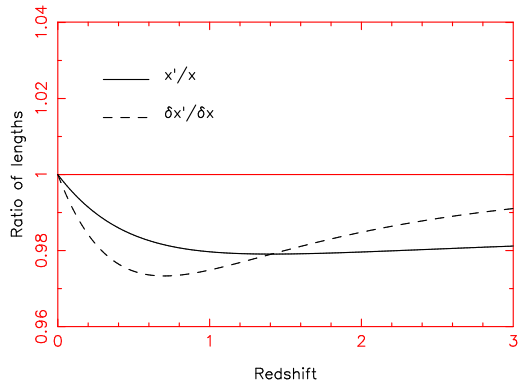}
\caption{The length distortion of a rod as a function of redshift,
supposing the true cosmology is $\om = 0.3$, $w_{true} = -1$ and the assumed
cosmology is $\om ' = 0.3$, $w_{ass} = -0.9$.  The dashed and solid lines
illustrate respectively the distortion if the rod is oriented radially
(i.e. $dx'/dx$) and tangentially (i.e. $x' \, d\theta/x \, d\theta =
x'/x$).  Thus it can be seen that the primary effect of assuming an
incorrect value of $w$ is a re-scaling of distances away from their
true values.}
\label{figdistort}
\end{figure}

Suppose $\om = 0.3$ and the true dark energy parameter is $w_{true} = -1$.
Consider a measuring rod located at $z = 1$.  If the assumed cosmology
is $w_{ass}= -0.9$ (but constant with redshift), then using equation \ref{eqcosmo} the length
distortion of the rod is $dx'/dx = 0.975$ if oriented radially
(i.e. its ends at fixed redshifts) and $x'/x = 0.980$ if oriented
tangentially (i.e. its ends at fixed angular positions).  Thus the
zeroth order effect is a re-scaling in the length of the rod by
$\simeq2$\% and the first order effect is a radial/transverse shear.
The full redshift dependence of these quantities is illustrated by
Figure \ref{figdistort}.

\placefigure{figdistort}

This distortion of spatial scales carries over directly into
$k$-space, and suggests that a measurement of $k_A$ with 2 per cent
precision at $z \sim 1$ translates into a measurement of $w$ with an
accuracy $\Delta w \approx 0.1$.  The corresponding figures at $z=3$
are $dx'/dx = 0.991$ and $x'/x = 0.981$; the smaller distortion in
$dx$ suggests that the method is slightly less powerful when applied
at $z \sim 3$.

We note that in detail, when combined with CMB observations, the
assumption of constant $\om$ in this comparison is not quite right.
The CMB power spectrum accurately constrains the angular diameter
distance to the surface of last scattering, which removes one degree
of freedom in the parameters: we cannot vary $\om$ and a constant $w$
completely independently.  In addition, observations by the Planck
satellite will accurately determine the quantity $\om h^2$ within a
few years.  However, in this initial treatment, we consider our
experiment as independent and suppose that $\om$ is specified by
external datasets.

It is interesting to contrast the baryonic oscillations method
presented here with the Alcock-Paczynski test.  The latter does not
employ a standard length-scale known in advance, but rather compares
quantities parallel and perpendicular to the line-of-sight.
Physically, it is sensitive to distortions in $dx/x$.  Numerically,
this amounts to the {\it difference} between the solid and dashed
lines in Figure \ref{figdistort} ($\frac{dx'/x'}{dx/x} - 1 \approx
\frac{dx'}{dx} - \frac{x'}{x}$ for small distortions).  The baryon
wiggles method uses a known ruler and is therefore independently
sensitive to distortions in $x$ and $dx$, through the parallel and
perpendicular components of the power spectrum (although in this
initial treatment we average the power spectrum over angles).
Numerically, this is equivalent to the {\it absolute deviation} from
unity of the solid and dashed lines in Figure \ref{figdistort}.  Thus
the ``lever arm'' for detecting any departures from $w = -1$ is a
factor $\sim 4$ higher at $z = 1$ for our method and, in contrast to
the Alcock-Paczynski test, it does not depend on small-scale
non-linear clustering details.

\subsection{Simulations}

We simulated surveys in the redshift ranges $0.5 < z < 1.3$ and $2.5 <
z < 3.5$, as before, using the same measured ranges of $k$ and linear
bias factors.  In each case we defined the survey volume by a
$20^\circ \times 20^\circ$ angular patch.  This generated a volume $V
= 3.9 \, \vs$ for the $z \sim 1$ survey and $V = 8.0 \, \vs$ at $z
\sim 3$.  We assumed redshift catalogues of $N = 10^6$ galaxies at $z
\sim 1$ and $N = 5 \times 10^5$ galaxies at $z \sim 3$, such that the
measurements were not limited by shot noise.  These examples are
somewhat arbitrarily chosen, but are indicative of future large-scale
redshift surveys.  We selected the galaxies from a redshift
distribution that populated the survey volume approximately uniformly.

We allowed the baryon density to vary over the range $0.1 < \ob/\om <
0.2$ (centred on the result $\ob/\om = 0.15$ assumed in Section
\ref{seclammeas}).  For each corresponding power spectrum we generated
a set of galaxy realizations and converted these to redshift
catalogues supposing a ``true'' cosmology $w = -1$.  We then
re-measured $P(k)$ for each realization for a range of ``assumed''
cosmologies, $-1.3 < w < -0.7$.  There are some theoretical reasons to
suppose $w \ge -1$, but we do not impose this restriction on our
analysis (Caldwell 1999).

\begin{figure}[t]
\includegraphics[width=5cm,angle=-90]{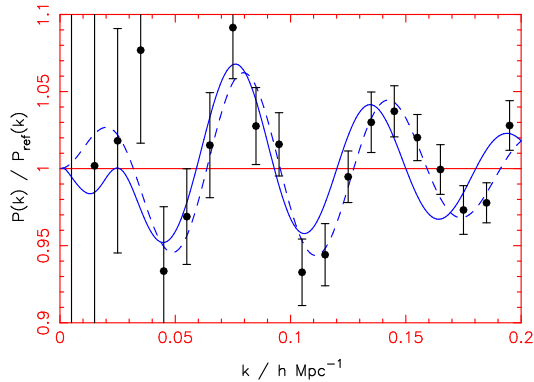}
\caption{Power spectrum measurement for a simulated survey with the
same parameters as Figure \ref{figpkreal}, except that the value
$w=-0.8$ has been incorrectly assumed.  The wavescale of the fitted
function (the dashed line) is spuriously distorted compared to the
theoretical expectation (the solid line).}
\label{figpkdist}
\end{figure}

For each assumed value of $w$ we used 400 different galaxy
realizations to derive a distribution of fitted wavescales in
$k$-space.  If the assumed value of $w$ differs from the ``true''
cosmology $w = -1$, then the fitted wavescales will be distorted (see
Figure \ref{figpkdist}) and their distribution will not be centred on
the expected wavescale (computed from Eisenstein \& Hu 1998, equation
26).  This allowed us to assign a likelihood to each value of $w$,
based on the position of the expected wavescale in the measured
distribution.  For example, 68 per cent of the fitted wavescales lie
between the $16^{\rm th}$ and $84^{\rm th}$ percentiles of the
measured distribution.  Thus if the expected wavescale lies at the
$16^{\rm th}$ percentile of the distribution, the value of $w$ is
rejected with ``$1\sigma$ significance''.  Likewise, if the expected
wavescale lies at the $2.5^{\rm th}$ percentile of the distribution,
the value of $w$ is rejected with ``$2\sigma$ significance''.

\placefigure{figpkdist}

\begin{figure}[t]
\includegraphics[width=5cm,angle=-90]{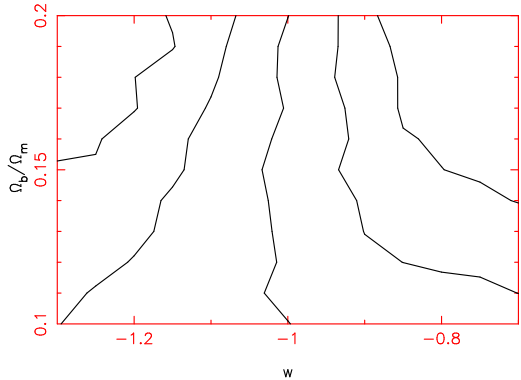}
\caption{Constraints on the dark energy parameter $w$ produced by
measuring the baryonic oscillations in the galaxy power spectrum, for
a potential large-scale redshift survey of 400 deg$^2$ at $z \sim 1$.
We assumed that the linear power spectrum can be successfully measured
for $k < 0.2 \, h$ Mpc$^{-1}$.  The survey parameters are $V = 3.9 \,
\vs$, $N = 10^6$, $b = 1$.  We assumed a ``true'' cosmology $w = -1$
and plot $0\sigma$, $1\sigma$ and $2\sigma$ confidence limits on the
measured value of $w$ for varying $\ob$.  These contours correspond to
the theoretical value of $k_A$ lying at the $2.5^{\rm th}$, $16^{\rm
th}$, $50^{\rm th}$, $84^{\rm th}$ and $97.5^{\rm th}$ percentiles of
the observed wavescale distribution.  The $50^{\rm th}$ percentile is
slightly displaced from $w = -1$ because the full Eisenstein \& Hu
(1998) transfer function is used in the simulations and the
approximate analytic formula (Eisenstein \& Hu 1998, equation 26) is
used for the theoretical prediction.}
\label{figwmeasz1}
\end{figure}

\begin{figure}[t]
\includegraphics[width=5cm,angle=-90]{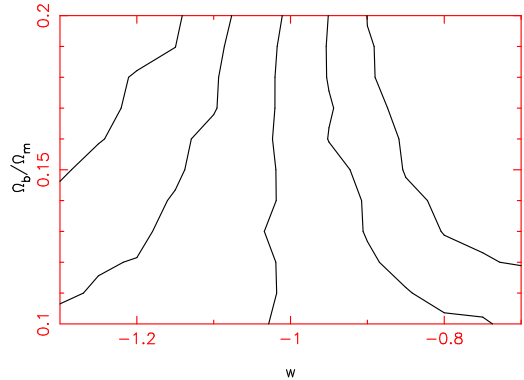}
\caption{The same plot as Figure \ref{figwmeasz1} for a simulated
survey of 400 deg$^2$ at $z \sim 3$, supposing that the linear power
spectrum can be successfully measured for $k < 0.3 \, h$ Mpc$^{-1}$.
The survey parameters are $V = 8.0 \, \vs$, $N = 5 \times 10^5$, $b =
3$.}
\label{figwmeasz3}
\end{figure}

Figures \ref{figwmeasz1} and \ref{figwmeasz3} display the final
results for the $z \sim 1$ and $z \sim 3$ surveys: $1\sigma$ and
$2\sigma$ confidence levels for the value of $w$ as a function of
$\ob$.  For the fiducial value of $\ob/\om = 0.15$ we can indeed
measure $w$ to a $1\sigma$ accuracy of $\pm 0.1$, bearing out our
back-of-the-envelope calculation in Section \ref{secwintro}.  This of
course assumes the fiducial model $w = -1$; the error bar will have
some dependence on the ``true'' value of $w$, which we do not
investigate here.

\placefigure{figwmeasz1}

\placefigure{figwmeasz3}

For comparison, we extended the angular area of the simulated $z \sim
1$ survey by a factor of five, generating $N = 5 \times 10^6$ galaxies
over a volume $V = 30 \, \vs$.  In this case, the wiggle wavescale
and dark energy parameter can be measured with respective accuracies
$\Delta k_A/k_A \approx 0.8$ per cent and $\Delta w \approx 0.04$
(assuming $\ob/\om = 0.15$).

\section{Discussion}
\label{secdiscuss}

Baryonic oscillations offer a measurement of the dark energy equation
of state that complements other methods.  The precision obtainable
from the surveys simulated above, $\Delta w \approx 0.1$ ($68$\%
confidence), is comparable to that achieved by combining together all
existing relevant observations.  The current high-redshift supernovae
data requires $w < -0.55$ at $95$\% confidence for a flat geometry
(Garnavich et al. 1998), and the latest (WMAP) CMB measurements, taken
together with other datasets, imply $w < -0.78$ at $95$\% confidence
(Spergel et al. 2003).  Moreover, if deviations are to be detected
from the $w = -1$ paradigm, observations are required at $z \sim 1
\rightarrow 3$ where we expect the corresponding cosmological effects
to be strongest: Universal expansion switches from deceleration to
acceleration at this epoch.

A high-quality sample of 2000 high-redshift supernovae distributed
from $z = 0.2$ to $z = 1.7$, provided by the SNAP satellite (Aldering
et al. 2002; {\tt http://snap.lbl.gov}), promises to measure a constant $w$ to a
precision $\Delta w = 0.05$ ($68$\% confidence) and constrain its time variation. 
The latter is further improved when combined with
the accurate measure of $\om h^2$ provided by Planck satellite
observations of the CMB (Frieman et al. 2003).  The baryon
oscillations and supernovae methods of probing the dark energy should
powerfully complement each other.  Supernovae provide a ``standard
candle'' approach (i.e. a measurement of the luminosity distance),
whereas baryon oscillations provide a ``standard rod'' approach,
measuring this rod in both the transverse and radial directions.  The
transverse rod measures the angular diameter distance.  In
conventional cosmology and physics, luminosity distance and angular
diameter distance are both related to $x(z)$ in equation \ref{eqcosmo}
by $(1+z)$ factors.  However, this is not true of all scenarios; for
an exotic alternative involving photons oscillating into undetectable
light axions see Cs{\' a}ki et al. (2002).  The baryonic oscillations
additionally measure the standard rod radially, yielding a measurement
of $dx/dz$ which is equivalent to the Hubble parameter $H(z)$.  To the
best of our knowledge the baryonic oscillations provide the only
direct route to $H(z)$ at high redshift.

The efficacy of the supernova approach is determined by how accurately
systematic errors can be controlled.  These systematic astrophysical
effects include the possible presence of non-standard host galaxy
extinction, the evolution of supernovae with redshift, the effect of
gravitational lensing, and variable Milky Way dust extinction.  It is
important to correct the supernova magnitudes by a light-curve factor
(Perlmutter et al. 1997) and/or colour/spectrum factor (Riess et
al. 1998).  In contrast, the acoustic oscillations method is not
limited by systematic errors, but by statistical uncertainties due to
sample size.

The two approaches also differ in their requirement of a local
calibration.  In the supernova method, cosmological parameters follow
from the relative brightnesses of supernovae at different redshifts:
hence the Nearby Supernova Factory ({\tt http://snfactory.lbl.gov})
plans a detailed study of several hundred low-redshift supernovae.  In
contrast, the measuring-rod length for the acoustic peaks is given
directly by fundamental physics ($\om h^2$ and $\ob h^2$) and does not
require additional observations at $z = 0$.

The use of counts of galaxies or clusters as probes of dark energy
also depends on understanding critical systematics such as the
variation with redshift of the intrinsic number density of the
objects, and the ability to identify systems of comparable mass at
different redshifts.

The approach presented here is essentially a geometrical method,
similar in conception to the Alcock-Paczynski test.  A classic
application of the Alcock-Paczynski test is to small-scale galaxy
clustering parallel and perpendicular to the line-of-sight (Ballinger
et al. 1996).  However, in this case it is very difficult to
distinguish the geometrical distortions imprinted by an incorrect
cosmology from redshift-space distortions.  In contrast, baryonic
oscillations probe clustering on much larger scales, where
redshift-space distortions are insignificant.

In our very empirical approach, the measurement of $w$ depends only on
knowledge of the positions of the acoustic peaks and troughs in the
power spectrum (and not on the detailed shape and amplitude of the
power spectrum).  It is the positions of these peaks, our ``standard
rod'', which are the least model-dependent aspects of the power
spectrum.  Detailed models exist for physics before recombination,
tested by increasingly accurate observations of the CMB power
spectrum.  The subsequent physics governing the growth of structures
lies firmly in the well-understood linear regime for the scales of
interest.

However, three important issues have been ignored in this initial
investigation and will be considered in a future study (see also Seo
\& Eisenstein, in prep.).  The most important unexplored effect in the
use of baryonic oscillations is the possible existence of more
complicated galaxy biassing schemes (a non-linear bias that may evolve
with redshift).  This said, linear scale-independent bias should not
be a bad approximation on large scales (e.g. Peacock \& Dodds 1994).
Secondly, we have restricted our attention to models with a constant,
redshift-independent value of $w$.  Our approach can be simply
extended to constrain models of varying $w$, such as the common
parameterizations $w(z) = w_0 + w_1 \, z$ or $w(z) = w_0 + w_a \, z/(1+z)$ 
which will be the subject of
a future study.  Thirdly, we have neglected the uncertainty in the
value of $\om$, which clouds precise knowledge of the expected
positions of the acoustic peaks.

It is easy to estimate the implications of uncertainty in $\om$ for
our method, assuming a ``true'' cosmology $\om = 0.3$, $w = -1$.  We
saw in Section \ref{secwintro} that a shift $\Delta w = 0.1$ produces
a radial distortion (in $dx/dz$) of $\approx 2$ per cent at $z = 1$.
Using equation \ref{eqcosmo}, a similar distortion is produced by a
shift $\Delta\om \approx 0.02$.  However, the sound horizon is also
changed by $\approx 1.5$ per cent in the same sense (using the fitting
formula of Eisenstein \& Hu 1998, equation 26), thus the effects
largely cancel out.  The ``overall'' radial distortion (relative to
the sound horizon) becomes 2 per cent when $\Delta\om \approx 0.07$,
which is reassuring: combining the WMAP CMB results with other
astronomical data sets produces a current measurement of $\om$ with
accuracy $\Delta\om \approx 0.04$ (Spergel et al. 2003), and this
situation should improve in the light of new data.  However, the $z
\sim 3$ survey demands much tighter knowledge of $\om$ than the $z
\sim 1$ survey, because the $(1+z)^3$ term inside the $\sqrt{\ }$ in
equation \ref{eqcosmo} starts heavily winning and the distortion due
to a shift in $w$ decreases.  The result is that at $z = 3$, the shift
$\Delta \om \approx 0.01$ produces the same overall radial distortion
as $\Delta w = 0.1$.  Thus the application of this method at $z = 3$
requires more accurate knowledge of $\om$ than is achievable from
current data.

It is worth considering whether an imaging survey using photometric
redshifts could accurately measure the acoustic peaks (see also
Eisenstein 2002).  Let us introduce a Gaussian scatter $\sigma_z =
(1+z) \sigma_0$ in each galaxy redshift.  This corresponds to a
smearing in co-moving co-ordinates of magnitude $\sigma_r = \sigma_z
(dx/dz)$.  Taking a concrete example, if $\sigma_0 = 0.03$ (a
realistic estimate of the best possible current accuracy of
photometric redshifts, e.g. Wolf et al. 2003) then $\sigma_r \approx
100 \, h^{-1}$ Mpc at $z=1$.  The effect on a spherically symmetric
$P(k)$ in ($k_x$,$k_y$,$k_z$)-space is a damping of all modes by a
factor $\exp{(-k_z^2 \sigma_r^2)}$ (in the flat-sky approximation),
where the $z$-axis is the line-of-sight.  Considering the modes in a
spherical shell of radius $k = 0.2 \, h$ Mpc$^{-1}$, the majority are
too heavily damped to provide signal.  The surviving useful modes
(with $|k_z| < 1/\sigma_r \approx 0.01 \, h$ Mpc$^{-1}$) are located
within a ``ring'' in $k$-space, corresponding to a reduction in the
number of useful modes by a factor $k \sigma_r \approx 20$.  To
achieve the same power spectrum precision on this scale $k$, the sky
area of the imaging survey (i.e. density of states in $k$-space) must
be increased by this same factor $\approx 20$ such that the ring
contains the same number of modes as the original spherical shell.

However, the most significant disadvantage of the photometric
redshifts approach is one loses the capacity to decompose the power
spectrum into its radial and tangential components, which separately
constrain the Hubble constant and angular diameter distance through
distortions in $dx$ and $x$ (Eisenstein 2002).  In our initial
``proof-of-concept'' treatment, we have averaged the power spectrum
over angles and not pursued these separate constraints, which
certainly merit more detailed calculation (see Seo \& Eisenstein, in
prep.).  As noted above, the radial component of $P(k)$ is damped by a
factor $\exp{(-k^2 \sigma_r^2)}$, which becomes unacceptably small
unless $\sigma_r \lesssim 1/k$ or (taking $k = 0.2 \, h$ Mpc$^{-1}$)
$\sigma_0 \lesssim 0.001$ at $z=1$.  Spectroscopic resolution $R \sim
1000$ is thus required for sufficiently accurate redshifts.

Finally, how realistic is it to carry out such redshift surveys in
practice?  Our simulated surveys involve $\sim 10^6$ galaxies over
$\sim 400$ deg$^2$ of sky.  The exposure times for spectra of these
objects would be of order 1--2 hours on 8-metre class telescopes,
based on existing data (e.g. Steidel et al. 1998).  An 8-metre
telescope with a $1^\circ$ diameter field of view, capable of taking
spectra of up to 3000 galaxies simultaneously, could accomplish such a
survey in 60--120 nights.  Such an instrument is eminently feasible
and has been proposed several times (Glazebrook 2002).  In particular,
a detailed concept design has been put together for the Gemini
telescopes (Barden 2002).  The observing time and data volume of such
a survey is very similar to that of the 2dF Galaxy Redshift Survey
(Colless et al. 2001).

\section{Conclusions}

This initial study has demonstrated that, under a simple set of
assumptions, baryonic oscillations in the galaxy power spectrum may be
used to measure accurately the equation of state of the dark energy.
For example, a survey of three times the Sloan volume at mean redshift
$z \sim 1$ can measure the sound horizon (i.e. the wiggle scale) with
an accuracy of 2 per cent, and the parameter $w$ (assumed constant)
with precision $\Delta w = 0.1$, assuming $\ob/\om = 0.15$.  This
probe of the dark energy is complementary to the supernova method,
with an entirely different set of uncertainties dominated by sample
size rather than systematic effects.  It would provide the second
independent pillar of evidence for the current epoch acceleration of
the Universal expansion. The constraints on $w$ become tighter as the
survey volume is enlarged.  We conclude that delineation of the
baryonic peaks, and their use as a standard cosmological ruler to
constrain cosmological parameters, should form an important scientific
motivation for the next generation of galaxy redshift surveys.

\acknowledgments

We thank Warrick Couch, Daniel Eisenstein, Louise Griffiths, Charley
Lineweaver, John Peacock and Alex Szalay for useful comments on
earlier drafts of the manuscript.  Karl Glazebrook and Chris Blake
acknowledge generous funding from the David and Lucille Packard
foundation and the Center for Astrophysical Sciences, Johns Hopkins
University.  Chris Blake acknowledges travel support from the
University of New South Wales.

\end{document}